\documentclass[twocolumn,showpacs,amsmath,amssymb,aps,floatfix]{revtex4}

\usepackage{graphicx}
\usepackage{dcolumn}
\usepackage{bm}

\begin{document}

\preprint{paperPA19 - CHI transf}

\title{Modeling long-range memory with stationary Markovian processes}        

\author{Salvatore Miccich\`e}
\affiliation{Dipartimento di Fisica e Tecnologie Relative, Universit\`a degli Studi di Palermo, \\
             Viale delle Scienze, Ed. 18, I-90128 Palermo, Italy}

\begin{abstract} 

\noindent In this paper we give explicit examples of power-law correlated stationary Markovian processes $y(t)$ where the stationary pdf shows tails which are gaussian or exponential. These processes are obtained by simply performing a coordinate transformation of a specific power--law correlated additive process $x(t)$, already known in the literature, whose pdf shows power-law tails $x^{-\alpha}$. We give analytical and numerical evidence that although the new processes (i) are Markovian and (ii) have gaussian or exponential tails their autocorrelation function still shows a power-law decay $\langle y(t) y(t+\tau) \rangle \propto \tau^{-\beta}$ where $\beta$ grows with $\alpha$ with a law which is compatible with  $\beta=\alpha/2-\eta$. When $\alpha<2(1+\eta)$ the process $y(t)$, although Markovian, is long-range correlated. Our results help in clarifying that even in the context of Markovian processes long-range dependencies are not necessarily associated to the occurrence of extreme events. Moreover, our results can be relevant in the modeling of complex systems with long memory. In fact, we provide simple processes associated to Langevin equations thus showing that long--memory effects can be modeled in the context of continuous time stationary Markovian processes.

\end{abstract}


\date{\today}

\maketitle

\section{Introduction}   \label{intro}

Stochastic processes have been used to model a great variety of systems in disciplines as disparate as physics \cite{BM,VanKampen81,risken,gardiner, Schuss,Oksendal}, 
genomics \cite{Waterman,Durbin}, finance \cite{Bouchaud,Mantegna}, climatology \cite{vanStorch} and social sciences \cite{Helbing}. 

One possible classification of stochastic processes takes into account the properties of their
conditional probability densities. In this respect, 
Markov processes play a central role in the modeling of 
natural phenomena. In the framework of discrete time stochastic processes, a process $x(t)$ is said to be a Markov process
if the conditional probability density $P(x_{n+1},t_{n+1}|x_{n},t_{n}; \dots ;x_{1},t_{1})$ depends only on the
last value $x_n$ at $t_n$ and not on the previous values $x_{n-1}$ at $t_{n-1}$,
$x_{n-2}$ at $t_{n-2}$, etc. More generally, the transition probability of any Markov
process fulfills the Chapman-Kolmogorov equation \cite{VanKampen81}.
It is worth noting that  a Markov process is fully determined 
by the knowledge of the probability density function (pdf)
$W(x,t)$ of the process and the transition probability
$P(x_{n+1},t_{n+1}|x_{n},t_{n})$. When the Markovian process is continuous
both in space and time, the time evolution of the pdf is described by a Fokker-Planck (FP)
equation. Such level of simplicity
is rather unique among stochastic processes. In fact, a 
non-Markovian process is characterized by an infinite hierarchy of 
transition probabilities. In this case, the time evolution of the pdf is described by a Master Equation rather than a simpler FP equation. 

Another classification of stochastic processes 
considers the nature of correlation
of the random variable. Under this classification,
random variables are divided in short-range and 
long-range correlated variables. Short-range 
correlated variables are characterized by a finite mean of time-scales 
of the process whereas a similar mean time-scale 
does not exist for long-range correlated variables \cite{Beran94}. 
An equivalent definition can be 
given by considering the finiteness or infiniteness of the 
integral of the autocorrelation function of the random process
\cite{Cassandro78,Bouchaud90,Samorodnitsky94}.
In the presence of long-range correlation, the time integral $s(t)$ of 
the process $x(t)$ is a superdiffusive 
stochastic process showing 
$\langle |\Delta s(t)|^2\rangle \sim D_\gamma\,t^{\gamma}$ where $\gamma>1$ and $D_\gamma$ is a constant.
Superdiffusive stochastic processes have been observed in
several physical systems. A classical example 
is Richardson's observation that the relative separation $\ell$ of two particles moving
in a turbulent fluid at time $t$ follows the relation $<\ell^2(t)> \propto t^3$ \cite{Richardson26}. 
Other examples include anomalous kinetic in chaotic dynamics 
due to flights and trapping \cite{Geisel85}, dynamics
of aggregate of amphiphilic molecules \cite{Ott90}, 
dynamics of a tracer in a two-dimensional rotating flow 
\cite{Solomon93}, non-coding regions of complete genomes \cite{sciortino} and volatility in financial markets \cite{vola}.
  
Several stationary Markovian processes are short-range correlated. In fact, the paradigmatic Markovian process is the Ornstein--Uhlembeck (OU) one \cite{ornstein}, whose autocorrelation function is the exponential function $\rho(\tau) = e^{- \tau/T}$ where $T$ is the time-scale of the process. Although in the OU process there is one single time-scale, a general Markovian stationary process can be multi-scale, i.e. it may admit either a discrete or a continuum set of time-scales. In the last case, when the largest time-scale is removed to infinity the process can even be  long-range correlated. The paradigmatic Markovian process with power-law autocorrelation function is given by the family of processes considered in Ref. \cite{Zoller}. These are stationary Markovian power-law correlated processes that were introduced in the context of diffusion in optical lattices and semiclassically describe the motion of atoms in a one-dimensional optical lattice formed by two counterpropagating laser beams perpendicularly polarized. For a certain choice of the relevant paramenters the processes become long--range correlated. 

The existence of a power-law decaying autocorrelation function in the processes of Ref. \cite{Zoller} is intimately related to the existence of power-law tails in the stationary pdf. This is easily understood by considering that the processes of Ref. \cite{Zoller} describe particles moving in a confining Smoluchowski potential which asymptotically grows like $\log(x)$. If one compare such slow growth with the one associated to the OU process, whose Smoluchowski potential grows like $x^2$, it is easy to recognize that in the case of Ref. \cite{Zoller} (i) a particle can reach positions far away from the center of the potential because it is subject to a relatively weaker force and (ii) if a particle reaches a position $X$, then it is not suddenly recalled towards the center of the potential and therefore it can explore for relatively long times the regions around $X$. Loosely speaking, the time-series of the processes of Ref. \cite{Zoller} can show persistencies and clustering of extreme events. Such processes perfectly fit the features of the model proposed in Ref. \cite{Havlin}, where long-range dependencies are shown to explain the clustering of extreme events. However, one could have in principle slowly decaying autocorrelation functions without necessarily observing the occurrence of extreme events. One such example is given by the Fractional Brownian motion (FBm) \cite{Beran94} which is a stochastic process where the autocorrelation function decays like a power-law and the stationary pdf is gaussian. In this paper, in the context of Markovian processes, we give explicit examples of power-law correlated stationary processes where the stationary pdf shows tails which are gaussian or exponential. We will introduce such processes starting from appropriate coordinate transformations of an additive processes introduced in Ref. \cite{Zoller}. 

The paper is organized as follow. In section \ref{meto} we review the eigenfunction methodology used to analyze the correlation properties of a given stochastic process and introduce a specific power-law correlated process with power-law tails. In section \ref{gauss} and \ref{exp} we will present examples of power-law correlated stochastic processes with gaussian and exponentail tails in the stationary pdf respectively. In section \ref{concl} we will draw our conclusions.

\section{Power-law Tails in the pdf}   \label{meto}

In this section we will briefly review the family of stochastic processes introduced in Ref. \cite{Zoller} and whose ergodicity properties have been investigated in Ref. \cite{Lutz}. A similar class of such processes have been considered in Ref. \cite{lillo}.

Let us consider a continuous Markovian stochastic process $x(t)$ whose pdf $W(x,t)$ is described by
the FP equation with constant diffusion coefficient
$\partial_t W=-\partial_x(D^{(1)}(x)W)+D\,\partial^2_x W$. For the sake of simplicity, in this study we set $D=1$.
In general, the eigenvalue spectrum of the FP equation describing a stationary process
consists of a discrete part 
${\lambda_0=0,\lambda_1,...,\lambda_p}$ and a continuous part 
$]\lambda_c,+\infty[$  ($\lambda_c \geqslant \lambda_p$) 
associated with eigenfunctions $\varphi_{\lambda}$. The stationary
pdf is $W(x)=\varphi_{0}$. The FP equation 
with constant diffusion coefficient can be transformed into a Schr\"odinger 
equation \cite{risken} with a quantum potential 
$V_S(x)=(D^{(1)}(x))^2/4+\partial_x D^{(1)}(x)/2$.
The eigenvalue spectrum of the Schr\"odinger 
equation is equal to the eigenvalue spectrum of the
FP equation. The relation between the 
eigenfunctions of the FP equation and the 
eigenfunctions $\psi_{\lambda}$ of the Schr\"odinger equation
is $\varphi_{\lambda}=\psi_{\lambda}\psi_0$. For a stationary process the 2-point probability density function 
$W_2(x,t;x',t+\tau)$ can be expressed in terms of 
the eigenfunctions of the Schr\"odinger equation. Specifically, one can write
\begin{eqnarray}
           &&   \hspace{-0.15 in}
                W_2(x,t;x',t+\tau)~=~\psi_0(x)~\psi_0(x') \times \label{W2} \\
           &&   \hspace{-0.1 in}\left( 
                                    \sum_{\lambda=\lambda_1}^{\lambda_p}\,\psi_\lambda(x)\,\psi_\lambda(x')\,e^{- \lambda \tau}\! + \! 
                                    \int_{\lambda_c}^{+\infty} \! d \lambda\,\psi_\lambda(x)\,\psi_\lambda(x')\,e^{- \lambda \tau}
                                \right) .      \nonumber
\end{eqnarray}
Eq. $(\ref{W2})$ extends the analogous expression valid for a FP equation with only discrete spectrum \cite{risken} to the case in which there also exists a continuous part of the spectrum. By direct inspection, it can be shown that $W_2$ fulfills the Chapman--Kolmogorov equation. 
In order to evaluate the autocorrelation function $\rho(\tau)= (\langle x(t+\tau)x(t)\rangle-\langle x(t)\rangle^2)/(\langle x^2(t)\rangle-\langle x(t)\rangle^2)$ of the stochastic variable $x(t)$, we make use of the expression 
\begin{equation} 
                \langle x(t+\tau)x(t)\rangle=\sum_{\lambda=\lambda_1}^{\lambda_p}~C^2_{\lambda}e^{-\lambda \tau}+
                                             \int_{\lambda_c}^{+\infty}~C^2_{\lambda}e^{-\lambda \tau} d\lambda, \label{COV}
\end{equation}
where $C_{\lambda} \equiv \int dx~x\,\varphi_{\lambda}(x)$. Eq. $(\ref{COV})$ follows from Eq. $(\ref{W2})$ and from the definition of the autocovariance function
\begin{eqnarray} 
                \langle x(t+\tau)x(t)\rangle = \iint_{-\infty}^{+\infty} \! dx'\,dx \,
                                               x'\,x\,W_2(x,t;x',t+\tau).  \label{COValt}
\end{eqnarray}
Eq. $(\ref{COV})$ holds true under the assumption that the integrations in $\int dx'\,\int dx$ and $\int d\lambda$ can be interchanged. 

The asymptotic temporal dependence of the autocorrelation function can have a different behavior conditioned by the
properties of the eigenvalue spectrum \cite{GS1,SK,GS2,HL}. Specifically, following \cite{FAR} one can distinguish three different cases, depending on the existence of a continuum spectrum of eigenvalues and whether or not such spectrum is attached to the ground state.

In fact, the class of processes introduced in Ref. \cite{Zoller} belongs to the one admitting a continuum part of the spectrum attached to the ground state. In this paper we will consider the specific stationary Markovian processes associated with a quantum potential $V_S$ given by
\begin{eqnarray}
              V_S(x)=\left \{\begin{array}{ccc}
                                   -V_0~    &~{\rm{if}}~&~|x| \leqslant L, \\
                                            &           &                  \\
                                   V_1/x^2 ~&~{\rm{if}}~&~|x| >   L,
                   \end{array} \right.  \label{VSchimera}
\end{eqnarray} 
where $L$, $V_0$ and $V_1$ are positive constants. The reason for considering such specific potential, among all those fullfilling the requirements of Ref. \cite{Zoller}, is that it is exactly solvable and therefore it will allow us to perform most calculations analytically. 

The parameters $L$, $V_0$ and $V_1$ can be chosen in such a way that the spectrum contains one single discrete eigenvalue $\lambda_0=0$ and a continuous part for $\lambda>0$. As a result, the parameters $L$, $V_0$ and $V_1$ are not independent. In fact, the continuity of $\partial_x \psi_0$ in $x=L$ provides a relation between them. The Langevin equation of the process is
\begin{eqnarray}
          &&  \dot{x}=h(x)+ dz \nonumber \\ 
          &&  h(x)=\left \{ \begin{array}{cc}
               -2 \sqrt{V_0} \tan (\sqrt{V_0}x) &{\rm{if}}~~|x| \leqslant L ,\\
                 &   \\
                 (1-\sqrt{1+4~V_1})/ x   &{\rm{if}}~~|x| >   L .   
                                  \end{array} \right. \label{D1chimera} \\
          &&  V_1=L \tan\bigl( \sqrt{V_0} L\bigl) \Bigl(1+L \tan\bigl( \sqrt{V_0} L\bigl)\Bigl)   \nonumber                 
\end{eqnarray}
The associated FP equation describes the dynamics of an overdamped particle moving in a Smoluchowski potential $U(x)=- \int dx\,h(x)$ that increases logarithmically in $x$. For $|x| \leqslant L$, the eigenfunction of the ground state is
$\psi_0=B_0 \,\cos(\sqrt{V_0}\,x)$ whereas for $|x|>L$ it decays according to $\psi_0=A_0 \,x^{(1-\sqrt{1+4\,V_1})/2}$.
The constants $A_0$ and $B_0$ are set by imposing that $\psi_0$ is normalized and continuous in $x=L$. It is worth noting that for $|x|>L$ the stationary pdf $W(x)$ of the stochastic process is a power-law function decaying as $|x|^{-\alpha}$ 
with $\alpha=\sqrt{1+4\,V_1}-1$. The normalizability of the eigenfunction of the ground state is ensured if $\alpha>1$.
In the present study we consider stochastic processes with finite variance which implies $\alpha > 3$.
Due to parity arguments, only the odd eigenfunctions $\psi_\lambda^{(odd)}$ of the continuous spectrum give a non-vanishing contribution to $C_\lambda$. For $|x| > L$ the eigenfunction $\psi_\lambda^{(odd)}$ is a linear combination of Bessel functions $\psi_\lambda^{(odd)}=A_\lambda\,\sqrt{x} J_\nu(\sqrt{\lambda}\,x)+B_\lambda\,\sqrt{x}\, Y_\nu(\sqrt{\lambda}\,x)$ where $\nu=(\alpha+1)/2$. For $|x| \leqslant L$ we find $\psi_\lambda^{(odd)}=D_\lambda\,\sin(\sqrt{V_0+\lambda}\,x)$. The coefficients $A_\lambda$, $B_\lambda$ and $D_\lambda$ are fixed by imposing that $\psi_\lambda^{(odd)}$ and its first derivative are continuous in $x=L$ and that $\psi_\lambda^{(odd)}$ are orthonormalized with a $\delta$-function of the energy. Similar conditions apply to the even solutions.

By using these eigenfunctions we obtain an exact expression for $C_\lambda$. The further integration required in Eq. $(\ref{COV})$ to obtain $\langle x(t+\tau)\,x(t) \rangle$ cannot be performed analytically. By using Watson's lemma \cite{Olver74} and by considering that the first term of the Taylor expansion of $C_{\lambda}^2$ is proportional to $\lambda^{(\alpha-5)/2}$ for small values of $\lambda$, for large values of $\tau$ one gets $\langle x(t+\tau)\,x(t) \rangle \propto \tau^{-\beta}$ where $\beta=(\alpha-3)/2$. That indicates that this stochastic process is stationary, Markovian and asymptotically power-law autocorrelated. When $3<\alpha<4$ the process is long--range correlated.

\section{Gaussian Tails in the pdf}   \label{gauss}

\begin{figure} 
\begin{center}
              \includegraphics[scale=0.48] {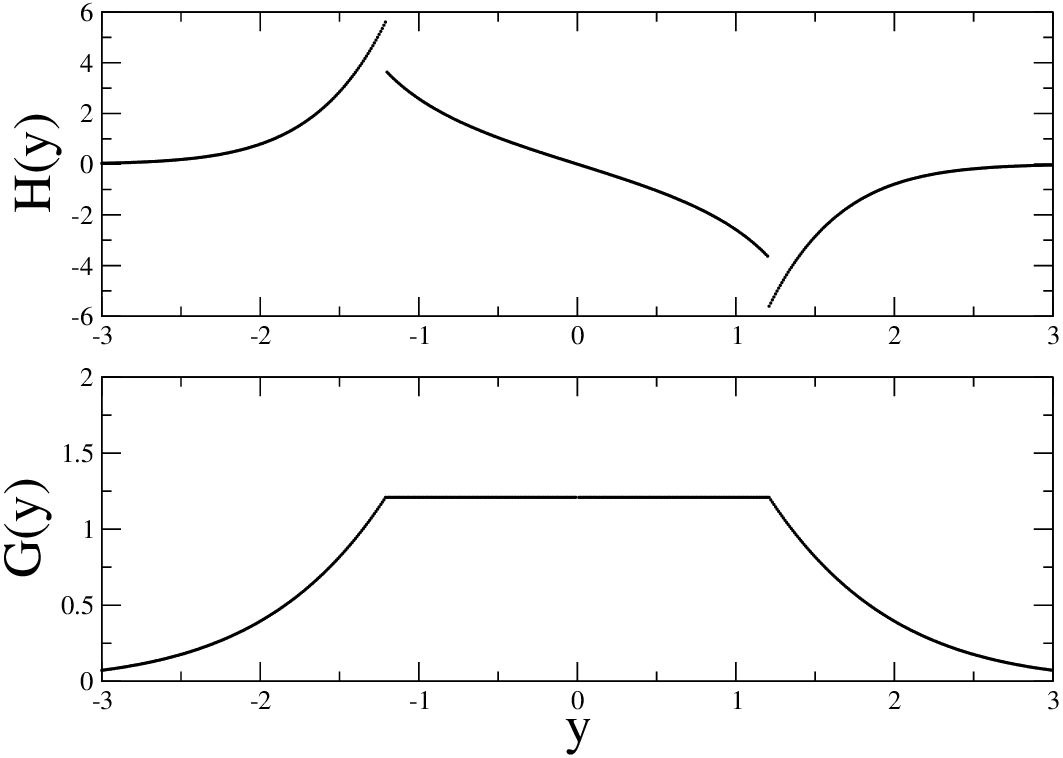} 
              \caption{The Figure shows the drift coefficient $H(y)$ (top panel) and diffusion coefficient $G(y)$ (bottom panel) of the process defined by the coordinate transformation of Eq. $(\ref{two_reg_pdf_gauss_transf})$ for the case when $L=1.0$, $V_0=0.987$ (i.e. $\alpha=3.05$) and $s=1.0$ (i.e. $r=1.2096$).}  \label{DriftDiffGauss}
\end{center}   
\end{figure}

In this section we explicitely present a stationary Markovian process with power-law decaying autocorrelation function and a stationary pdf with gaussian tails. In fact, let us consider the coordinate transformation:
\begin{eqnarray}
          &&    f_g(x)= \left \{ \begin{array}{l}
                                \sqrt{2\,s}~{\rm{Erf}}^{^{-1}}
                                       \Bigl[(1-r(x))\,{\rm{Erf}}(L r /\sqrt{2 s})+                         \\
                             \hspace{1.9 cm}     a(x)(1+\nu) \Bigl]     \hfill           ~|x| > L \\
                                                                                                   \\   
                                r\,x                                     \hfill           ~|x| \le L   
                       \end{array} \right.   \nonumber  \\
          &&    \nu={
                      {L\,r\,\sqrt{V_0}\,{\rm{sec}}(L\,\sqrt{V_0})^2\,+r\,\tan(L\,\sqrt{V_0})}
                      \over
                      {\sqrt{2\,\pi\,s\,V_0}~e^{L^2 r^2/2 s}}
                      }     \nonumber  \\
          &&    r(x)={
                      {2 A_0^2 (L x^\alpha-x L^\alpha)}
                      \over
                      {(\alpha-1)~L^{^\alpha}~x^{^\alpha}}
                     } \label{two_reg_pdf_gauss_transf}    
\end{eqnarray}
where $r$ is a real positive constant. By using the Ito lemma, one can show that, starting from the process of Eq. $(\ref{D1chimera})$, in the coordinate space $y=f_g(x)$ one gets a multiplicative stochastic process whose stationary pdf is:
\begin{eqnarray}
	     &&   W_g(y)= \left \{ \begin{array}{lll}
                                N_I~e^{- {1 \over{2~s}}y^2}                  & ~|y| > L~r  &  \\
                                                                             &             &  \\
                                N_{II}~\cos({\sqrt{V_0} y}/r)                & ~|y| \le L~r&  
                       \end{array} \right.   \label{two_reg_pdf_gauss_pdf} 
\end{eqnarray}
where $N_I $ and $N_{II}$ are normalization constants that can be analytically computed by imposing that $W_g(y)$ is continuous in $y=\pm L r$ and it is normalized to unity. The real constant $r$ is fixed by imposing that the diffusion coefficient $G(y)$ of the multiplicative stochastic process in the $y$ coordinate space is continuous in $y=\pm L r$. In Fig. $\ref{DriftDiffGauss}$ we show the drift coefficient $H(y)$ (top panel) and diffusion coefficient $G(y)$ (bottom panel) for the case when $L=1.0$, $V_0=0.987$ (i.e. $\alpha=3.05$) and $s=1.0$ (i.e. $r=1.2096$). The diffusion coefficient $G(y)$ is continuous in $y=\pm L r$ although its first derivative is discontinuous. The drift coefficient $H(y)$ suffers a discontinuity in $y=\pm L r$.  

The autocorrelation function of the process defined by the coordinate transformation of Eq. $(\ref{two_reg_pdf_gauss_transf})$ is given by $\rho_g(\tau)= (\langle y(t+\tau)y(t)\rangle-\langle y(t)\rangle^2)/(\langle y^2(t)\rangle-\langle y(t)\rangle^2)$ where:
\begin{eqnarray}
       &&   \langle y(t) y(t+\tau) \rangle=\int_0^\infty d \lambda\,{\cal{C}}_\lambda^{2} e^{- \lambda \tau} \nonumber \\
       &&   ~~ \label{Rgauss} \\
       &&   {\cal{C}}_\lambda=\int_{-\infty}^{+\infty} dx\,f_g(x)\,\psi_0(x)\,\psi_\lambda(x)\nonumber
\end{eqnarray}
where $\psi_0(x)$ and $\psi_\lambda(x)$ are the eigenfunctions of the process of Eq. $(\ref{D1chimera})$. Eq. $(\ref{Rgauss})$ can be used to numerically obtain the autocorrelation of the process defined by the coordinate transformation of Eq. $(\ref{two_reg_pdf_gauss_transf})$. 

\begin{figure} 
\begin{center}
              \includegraphics[scale=0.35] {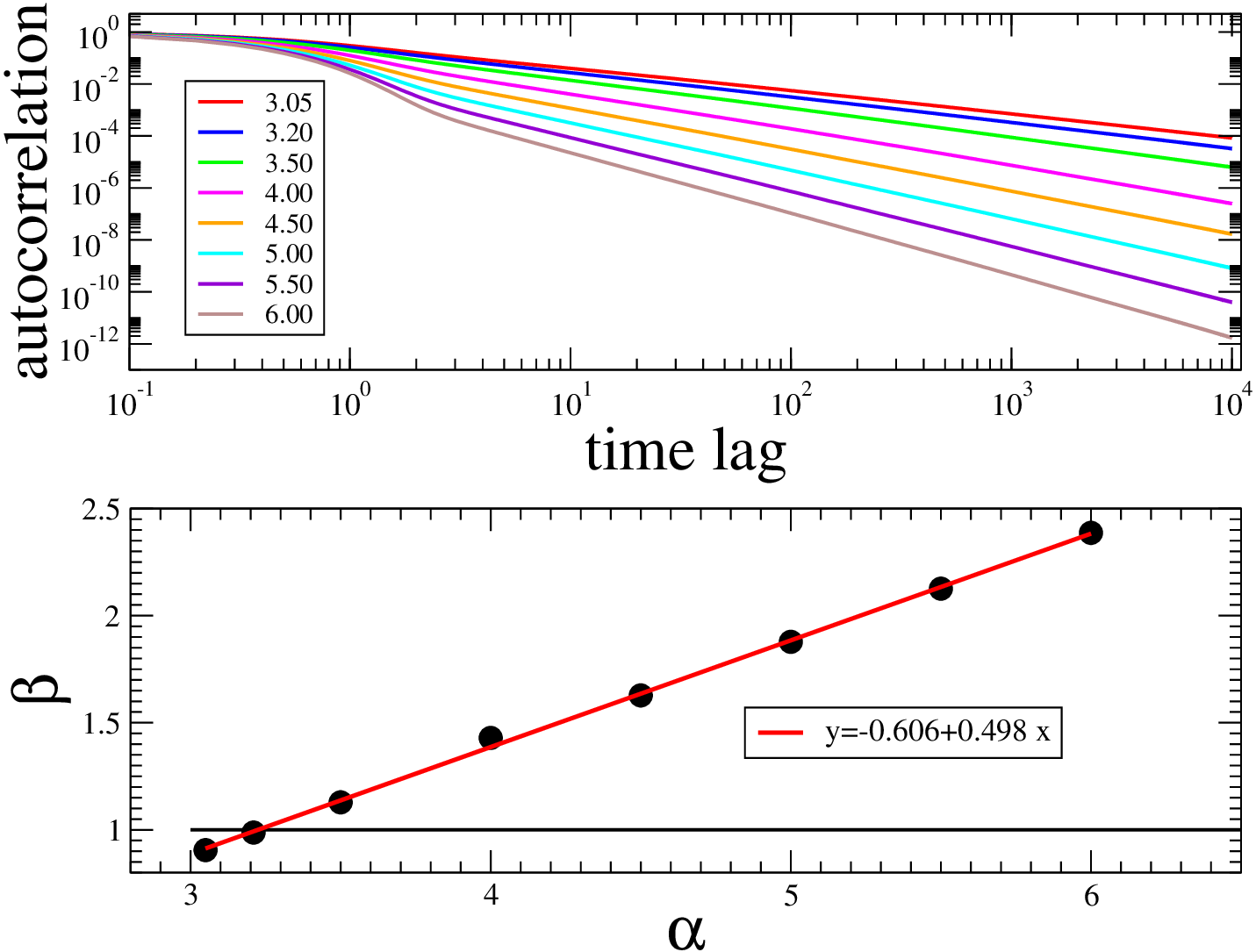} 
              \caption{In the top panel we report the results of the numerical integration of Eq. $(\ref{Rgauss})$ for the case when $L=1.0$, $s=1.0$ and the $V_0$ values are choosen in such a way that the parameter $\alpha$ assumes the values shown in the legend. In the bottom panel we report the values of the exponents $\beta_g$ obtained by performing a nonlinear fit of the autocorrelation function shown in the top panel.}  \label{fig:gaussall}
\end{center}   
\end{figure}

In the top panel of Fig. $\ref{fig:gaussall}$ we report the results of the numerical integration of Eq. $(\ref{Rgauss})$ for the case when $L=1.0$, $s=1.0$ and the $V_0$ values are choosen in such a way that the parameter $\alpha$ assumes the values shown in the legend. The asymptotic behaviour of these autocorrelation functions seems compatible with a power--law $\tau^{-\beta_g}$. In the bottom panel of Fig. $\ref{fig:gaussall}$ we report the values of the exponents $\beta_g$ obtained by performing a nonlinear fit of the autocorrelation function shown in the top panel. Such values show a dependence from the $\alpha$ parameter which seems compatible with a linear law $\beta_g=\alpha/2 - \eta_g$, with $\eta_g \approx 0.61$. For values $\alpha<2 \eta_g+1$ we get $\beta_g<1$, i.e. the stochastic process thus generated is long-range correlated.
It is worth noticing that the autocorrelation function $\langle y(t) y(t+\tau) \rangle$ does not show any dependance from the $s$ parameter.

\begin{figure} 
\begin{center}
              \includegraphics[scale=0.35] {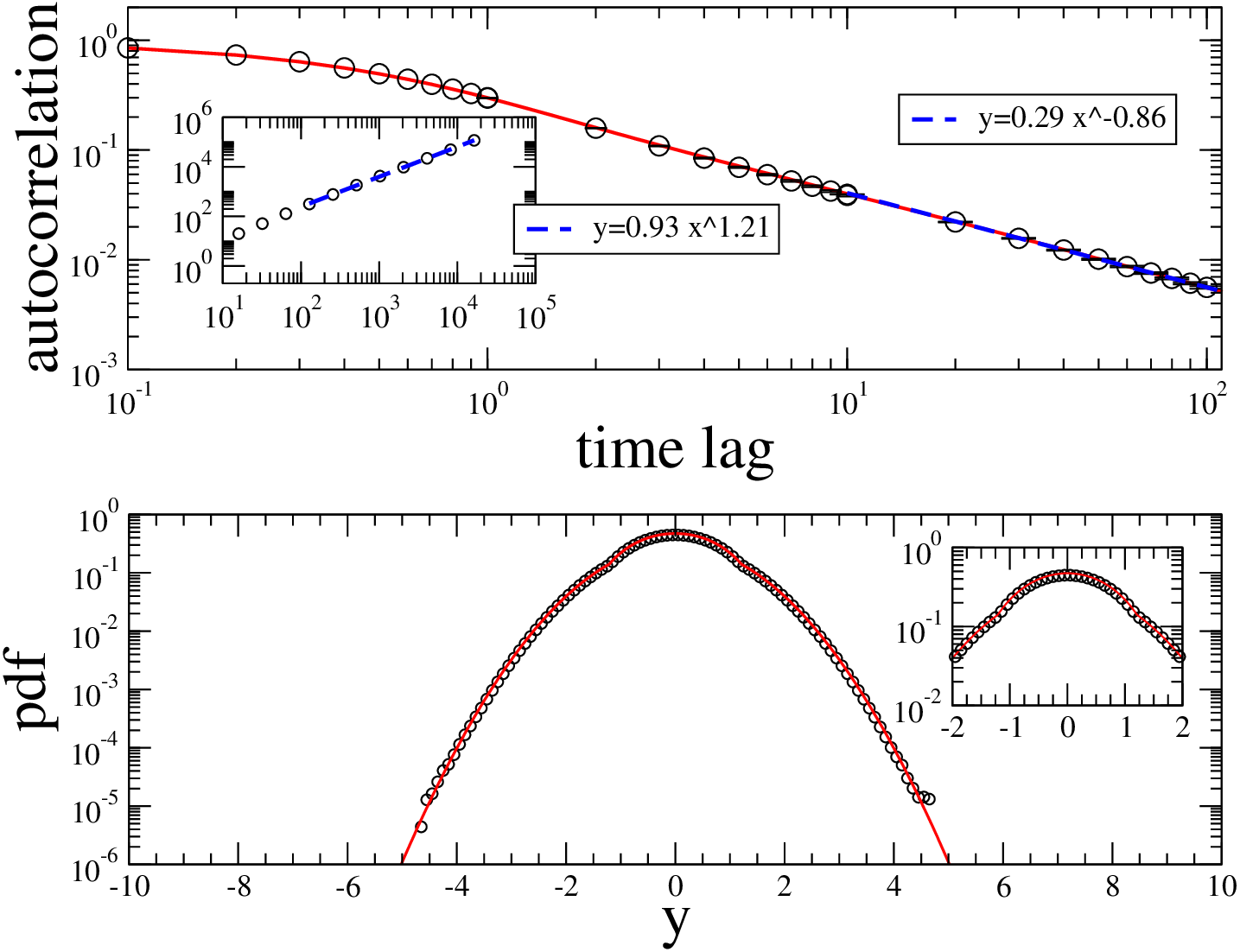} 
              \caption{The figure shows time--average numerical simulations performed according to Eq. $(\ref{ACtimeens})$ performed for the case when $L=1.0$, $V_0=0.987$ (i.e. $\alpha=3.05$) and $s=1.0$ (i.e. $r=1.2096$). The simulation parameters are: $M=10$ in the region $\tau \in [1,10]$ and $M=20$ in the region $\tau \geqslant 10$, $T=1.01\,10^{11}$. The time-step was $\Delta t=0.005$. In the top panel we show the results for the autocorrelation function. The solid (red) line shows the theoretical prediction obtained from Eq. $(\ref{Rgauss})$, while the open circles show the result of the numerical simulations. By performing a nonlinear fit (dashed blue line), the autocorrelation function shows an asymptotic decay compatible with a power-law $\tau^{-\beta_g}$, with $\beta_g=0.86$. In the inset of the top panel we show the numerical simulation (circles) relative to the mean square displacement $\langle |\Delta s(t)|^2\rangle$. A nonlinear fit (solid blue line) shows that $\langle |\Delta s(t)|^2\rangle \propto t^\delta$ with $\delta=1.21$. The bottom panel shows the stationary pdf of the process. Again the solid (red) line shows the theoretical prediction of Eq. $(\ref{two_reg_pdf_gauss_pdf})$, while the open circles show the result of the numerical simulations.}  \label{fig:gaussLR}
\end{center}   
\end{figure}

In the top panel of Fig. $\ref{fig:gaussLR}$ we show the results of numerical simulations of the autocorrelation function performed for the case when $L=1.0$, $V_0=0.987$ (i.e. $\alpha=3.05$) and $s=1.0$ (i.e. $r=1.2096$). 
The solid (red) line shows the theoretical prediction obtained from Eq. $(\ref{Rgauss})$, while the open circles show the result of the numerical simulations. By performing a nonlinear fit (dashed blue line), the autocorrelation function shows an asymptotic decay compatible with a power-law $\tau^{-\beta_g}$, with $\beta_g=0.86$. In the inset of the top panel we show the numerical simulation (circles) relative to the mean square displacement $\langle |\Delta s(t)|^2\rangle$ where $s(t)$ is the stochastic process obtained by integrating over time the process defined by the coordinate transformation of Eq. $(\ref{two_reg_pdf_gauss_transf})$. A nonlinear fit (solid blue line) shows that $\langle |\Delta s(t)|^2\rangle \propto t^\delta$ with $\delta=1.21$, thus confirming that we are observing a superdiffusive long-range correlated stochastic process. The bottom panel of Fig. $\ref{fig:gaussLR}$ shows the stationary pdf of the process. Again the solid (red) line shows the theoretical prediction of Eq. $(\ref{two_reg_pdf_gauss_pdf})$, while the open circles show the result of the numerical simulations. In the inset we show the same pdf in a shorter range of values in order to emphasize that inside the region $|y| \le L~r$ the pdf has a behaviour different from gaussian.

Those shown in Fig. $\ref{fig:gaussLR}$ are time--average numerical simulations performed according to the relation:
\begin{eqnarray}
              \rho(\tau)= {1 \over T} \int_0^T dt~x_*(t) x_*(t+\tau) \label{ACtime}
\end{eqnarray}
where $T$ is the length of the simulated time-series and $x_*(t)$ is one realization of the process. Indeed, in order to improve the statistical reliability of our numerical simulations, in the region $\tau\geqslant1$ we have also averaged over a number M of different realizations of the process:
\begin{eqnarray}
              \rho_T(\tau)= {1 \over M} \sum_{j=1}^M {1 \over T} \int_0^T dt~x_j(t)~x_j(t+\tau) \label{ACtimeens}
\end{eqnarray}
The data shown in the figure are the mean and the standard deviations of the $M$ autocorrelation values computed in each iteration for each time lag. The values of $M$ are $M=10$ in the region $\tau \in [1,10]$ and $M=20$ in the region $\tau \geqslant 10$. The size of each time-series was $T=1.01\,10^{11}$ with a time-step of $\Delta t=0.005$. The starting points of the simulated time-series were all the same with $x_j(0)=0.1$ where $j=1, \cdots, M$. In order to simulate the process in the $y$ coordinate space, we start by simulating the process of Eq. $(\ref{D1chimera})$ and compute $y=f(x)$ for each simulated $x$ value. However, we have explicitely checked that such procedure is equivalent to a direct simulation of the Langevin equation obtained starting from $H(y)$ and $G(y)$.

The existence of power-law correlated processes with gaussian tails does not contraddict the Doob Theorem \cite{Doob}. In fact, such theorem deals with the case when the process admits stationary pdfs and 2--point conditional transition probabilities which are both non singular and gaussian on the whole real axis.

\section{Exponential Tails in the pdf}   \label{exp}

In this section we explicitely present a stationary Markovian process with power-law decaying autocorrelation function and a stationary pdf with exponentially decaying tails. In fact, let us consider the coordinate transformation:
\begin{eqnarray}
                f_e(x)= \left \{ \begin{array}{lll}
                                {1 \over \gamma} \log \bigl(
                                                      \gamma (x-L) +e^{- \gamma L r}
                                                      \bigl) & ~|x| > L   &  \\
                                                             &            &  \\
                                r\,x                         & ~|x| \le L &  
                       \end{array} \right.   \label{two_reg_pdf_exp_transf} 
\end{eqnarray}
By using the Ito lemma, one can show that, starting from the process of Eq. $(\ref{D1chimera})$, in the coordinate space $y=f_e(x)$ one gets the multiplicative stochastic process described by:
\begin{eqnarray}
             &&   \dot{y}=H(y)+G(y)~\Gamma(t)  \label{two_reg_pdf_exp_proc} \\
	     &&   H(y)= \left \{ \begin{array}{lll}
                                - \gamma~e^{- 2 \gamma y}~{{(1+\alpha)\,e^{\gamma y}+ \Lambda}
                                                           \over
                                                           {e^{\gamma y} + \Lambda}}  & ~|y| > L\,r &  \\
                                                                                      &             &  \\
                                - 2 r \sqrt{V_0} \tan \bigl( \sqrt{V_0} y /r \bigl)   & ~|y| \le L\,r&  
                       \end{array} \right.   \nonumber \\
	     &&   G(y)= \left \{ \begin{array}{lll}
                                e^{- \gamma y}                                        & ~|y| > L\,r &  \\
                                                                                      &             &  \\
                                r                                                     & ~|y| \le L\,r  &  
                       \end{array} \right. \nonumber \\
             &&   \Lambda=\gamma L - e^{- \gamma L r}   \nonumber       
\end{eqnarray}
where $r$ is a real positive constant which is fixed by imposing that the diffusion coefficient $G(y)$ is continuous in $y=\pm L r$. It is straightforward to prove that such process admits the stationary pdf: 
\begin{eqnarray}
	     &&   W_e(y)= \left \{ \begin{array}{lll}
                                N_I~e^{- \gamma y}                       & ~|y| > L\,r &  \\
                                                                         &             &  \\
                                N_{II}~\cos \bigl( \sqrt{V_0} y /r \bigl)^2 & ~|y| \le L\,r&  
                       \end{array} \right.   \label{two_reg_pdf_exp_pdf} 
\end{eqnarray}
whose tails are exponential. $N_I $ and $N_{II}$ are normalizations constants that can be analytically computed by imposing that $W_e(y)$ is continuous in $y=\pm L r$ and it is normalized to unity.

The autocorrelation function of the process of Eq. $(\ref{two_reg_pdf_exp_proc})$ can be obtained starting from Eq.
$(\ref{Rgauss})$ with $f_g(x)$ now replaced by $f_e(x)$ of Eq. $(\ref{two_reg_pdf_exp_transf})$.

\begin{figure} 
\begin{center}
              \includegraphics[scale=0.35] {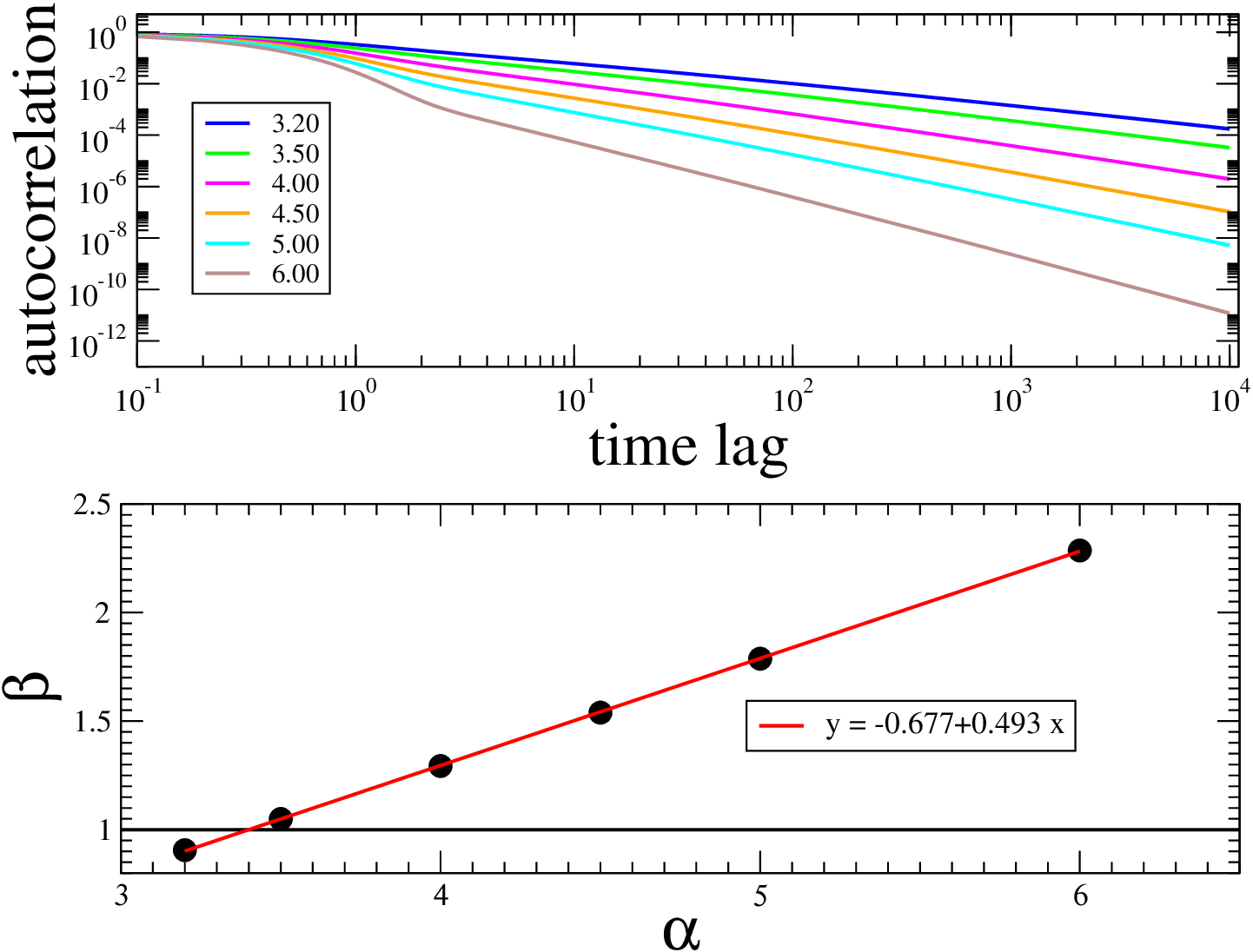} 
              \caption{In the top panel we report the results of the numerical integration of Eq. $(\ref{Rgauss})$ for the coordinate transformation of Eq. $(\ref{two_reg_pdf_exp_transf})$ when $L=1.0$, $\gamma=1.0$ and the $V_0$ values are choosen in such a way that the parameter $\alpha$ assumes the values shown in the legend. In the bottom panel we report the values of the exponents $\beta_e$ obtained by performing a nonlinear fit of the autocorrelation function shown in the top panel.}  \label{fig:expall}
\end{center}   
\end{figure}

In the top panel of Fig. $\ref{fig:expall}$ we report the results of the numerical integration of Eq. $(\ref{Rgauss})$ for the case when $L=1.0$, $\gamma=1.0$ and the $V_0$ values are choosen in such a way that the parameter $\alpha$ assumes the values shown in the legend. The asymptotic behaviour of these autocorrelation functions seems compatible with a power--law $\tau^{-\beta_e}$. In the bottom panel of Fig. $\ref{fig:expall}$ we report the values of the exponents $\beta_e$ obtained by performing a nonlinear fit of the autocorrelation function shown in the top panel. Such values show a dependence from the $\alpha$ parameter which seems compatible with a linear law $\beta_e=\alpha/2 - \eta_e$, with $\eta_e \approx 0.68$.
Differently form the gaussian case, now the autocorrelation function $\langle y(t) y(t+\tau) \rangle$ seems to show some dependance from the $\gamma$ parameter. As an example, we have computed the autocorrelation functions for the same values as above and with $\gamma=1.0$ replaced by $\gamma=10.0$. Again we find that $\eta$ depends upon $\alpha$ according to a linear law  $\beta_e=\alpha/2 - \eta_e$ where now $\eta_e \approx 0.62$.

\begin{figure} 
\begin{center}
              \includegraphics[scale=0.35] {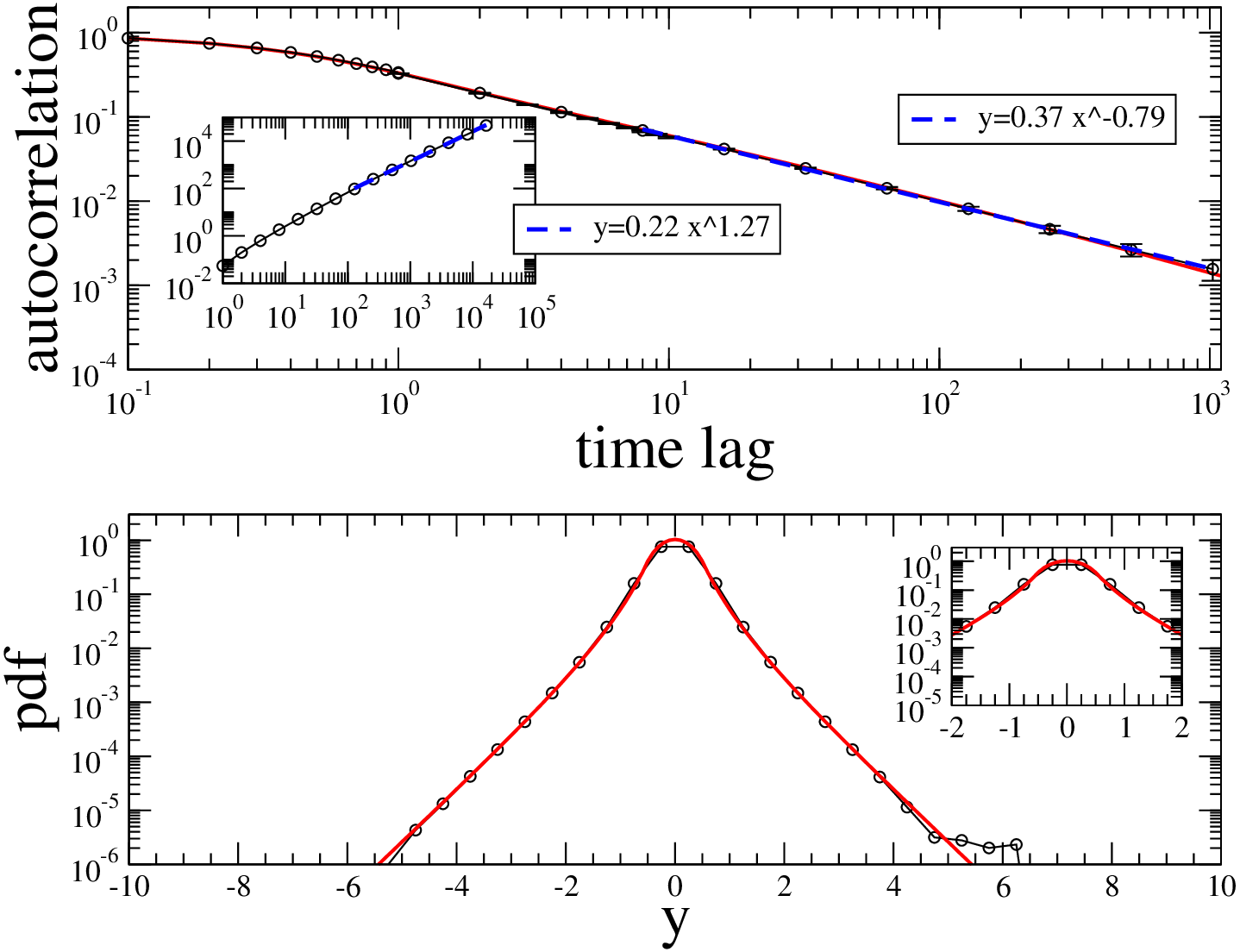} 
              \caption{The figure shows time--average numerical simulations performed according to Eq. $(\ref{ACtimeens})$ performed for the case when $L=1.0$, $V_0=1.020$ (i.e. $\alpha=3.21$) and $\gamma=1.0$ (i.e. $r=0.567$). The simulation parameters are: $M=10$ in the region $\tau \in [1,10]$ and $M=20$ in the region $\tau \geqslant 10$, $T=1.025\,10^{11}$. The time-step was $\Delta t=0.01$. In the top panel we show the results for the autocorrelation function. The solid (red) line shows the theoretical prediction obtained from Eq. $(\ref{Rgauss})$, while the open circles show the result of the numerical simulations. By performing a nonlinear fit (dashed blue line), the autocorrelation function shows an asymptotic decay compatible with a power-law $\tau^{-\beta_e}$, with $\beta_e=0.79$. In the inset of the top panel we show the numerical simulation (circles) relative to the mean square displacement $\langle |\Delta s(t)|^2\rangle$. A nonlinear fit (solid blue line) shows that $\langle |\Delta s(t)|^2\rangle \propto t^\delta$ with $\delta=1.26$. The bottom panel shows the stationary pdf of the process. Again the solid (red) line shows the theoretical prediction of Eq. $(\ref{two_reg_pdf_exp_pdf})$, while the open circles show the result of the numerical simulations.}  \label{fig:expLR}
\end{center}   
\end{figure}

In the top panel of Fig. $\ref{fig:expLR}$ we show the results for the case when $L=1.0$, $V_0=1.020$ (i.e. $\alpha=3.21$) and $\gamma=1.0$ (i.e. $r=0.567$). 
The solid (red) line shows the theoretical prediction obtained from Eq. $(\ref{Rgauss})$, while the open circles show the result of the numerical simulations. By performing a nonlinear fit (dashed blue line), the autocorrelation function shows an asymptotic decay compatible with a power-law $\tau^{-\beta_e}$, with $\beta_e=0.79$. In the inset of the top panel we show the numerical simulation (circles) relative to the mean square displacement $\langle |\Delta s(t)|^2\rangle$ where $s(t)$ is the stochastic process obtained by integrating over time the process defined by the coordinate transformation of Eq. $(\ref{two_reg_pdf_exp_transf})$. A nonlinear fit (solid blue line) shows that $\langle |\Delta s(t)|^2\rangle \propto t^\delta$ with $\delta=1.26$, thus confirming that we are observing a superdiffusive long-range correlated stochastic process. The bottom panel of Fig. $\ref{fig:expLR}$ shows the stationary pdf of the process. Again the solid (red) line shows the theoretical prediction of Eq. $(\ref{two_reg_pdf_exp_pdf})$, while the open circles show the result of the numerical simulations. In the inset we show the same pdf in a shorter range of values in order to emphasize that inside the region $|y| \le L~r$ the pdf has a behaviour different from exponential.

Those shown in Fig. $\ref{fig:expLR}$ are time--average numerical simulations performed according to Eq. $(\ref{ACtime})$. Differently from the previous case, when simulationg the process we directly consider the Langeving Equation of Eq. $(\ref{two_reg_pdf_exp_proc})$. Again, in order to improve the statistical reliability of our numerical simulations, in the region $\tau>1$ we have also averaged over a number M of different realizations of the process, according to Eq. $(\ref{ACtimeens})$. The data shown in the figure are the mean and the standard deviations of the $M$ autocorrelation values computed in each iteration for each time lag. The values of $M$ are $M=10$ in the region $\tau \in [1,10]$ and $M=20$ in the region $\tau \geqslant 10$. The size of each time-series was $T=1.025\,10^{11}$ with a time-step of $\Delta t=0.01$. The starting points of the simulated time-series were all the same with $x_j(0)=0.1$ where $j=1, \cdots, M$.

\section{Conclusions}   \label{concl}

In summary, we have shown new stationary Markovian processes which are power-law correlated and have a stationary pdf with tails that can be gaussian and exponential. The processes are obtained by simply performing a coordinate transformation of the additive process described in Eq. $(\ref{D1chimera})$. Starting from such specific process, we have given analytical evidence that the considered processes have the wanted stationary pdf and we have given numerical evidence that the autocorrelation function shows a power-law decay. Specifically, we find that for large values of time lag the autocorrelation function decays like $\langle y(t) y(t+\tau) \rangle \propto \tau^{-\beta}$ where $\beta$ grows with $\alpha$ with a law which seems compatible with $\beta=\alpha/2-\eta$ where $\eta$ is a parameter which depends from the specific tails of the stationary pdf. However, when the tails are gaussian $\eta$ does not show any dependance from the variance of the pdf. The above linear law holds true also in the case of the additive process of Eq. $(\ref{D1chimera})$, with $\eta=3/2$.

It is worth remarking that in principle more general processes can be obtained (i) by choosing different coordinate trasformations or (ii) by appropriately engineering the shape of the quantum potential $V_S(x)$ of Eq. $(\ref{D1chimera})$ in the region $[-L,L]$. This would result in a different shape of the stationary pdf in that region. When doing that, the asymptotic power-law behaviour of the autocorrelation function is not modified. In this paper we preferred to consider a linear transformation and $V_S(x)=-V_0$ in the region $[-L,L]$ only because this allows us to analytically obtain the eigenfunctions on the whole real axis and to obtain a numerical theoretical prediction for the autocorrelation function of the stochastic processes considered. 

Starting from the process of Eq. $(\ref{D1chimera})$, stationary pdfs with tails different from exponential or gaussian ones can be obtained by introducing appropriate coordinate transformations. In all cases the autocorrelation functions can be obtained, at least numerically, by using the same approach illustrated in this paper. 

To our knowledge, this is the first evidence of power-law correlated stationary Markovian processes with gaussian or exponential tails in the stationary pdf. It is worth remarking that the existence of power-law correlated processes with gaussian tails does not contraddict the Doob theorem \cite{Doob}, because the Doob theorem deals with the case when the process admits a stationary pdf and a 2--point conditional transition probability which are both gaussian on the whole real axis and non singular. In our case we only have gaussian tails in the stationary pdf.

Our results help in clarifying that even in the context of Markovian processes long-range dependencies are not necessarily associated to the occurrence of extreme events. It is worth mentioning that the processes introduced in section \ref{gauss} and section \ref{exp} are in the basin of attraction of the Gumbel distribution \cite{Embrechts}, although the one of Eq. $(\ref{D1chimera})$ is in the basin of attraction of the Frechet distribution. 

Moreover, our results can be relevant in the modeling of complex systems with long memory. In fact, processes with long-range interactions are often modeled by means of the Fractional Brownian motion (FBm), multifractal processes, memory kernels and other. Here we provide simple processes associated to Langevin equations thus showing that memory effects can still be modeled in the context of continuous time stationary Markovian processes, i.e. even assuming the validity of the Chapman-Kolmogorov equation.


%
%
%
%


\end{document}